%% file: paper.tex
\documentclass[conference]{IEEEtran}

\usepackage{breakcites}
\usepackage{url}
\usepackage{algorithm}
\usepackage{algorithmic}
\usepackage{amsmath}
\usepackage{amsthm}
\usepackage{graphicx}
\usepackage{subfigure}
\usepackage{multirow}
\usepackage{bm}
\usepackage{float}
\usepackage{mathtools}
\usepackage{balance}
\usepackage{fancyhdr}

\usepackage[colorlinks, linkcolor=red, anchorcolor=blue, citecolor=green]{hyperref}
\usepackage{cleveref}

\let\OLDthebibliography\thebibliography
\renewcommand\thebibliography[1]{
  \OLDthebibliography{#1}
  \setlength{\parskip}{0pt}
  \setlength{\itemsep}{0pt plus 0.3ex}
}

\crefname{section}{§}{§§}
\newtheorem{property}{Property}[section]
\newcommand\figcaption{\def\@captype{figure}\caption}
\newcommand\tabcaption{\def\@captype{table}\caption}
\begin{document}

\title{DaI: Decrypt and Infer the Quality of Real-Time Video Streaming}

\author{Sheng Cheng}

\maketitle


\input{body}

\bibliographystyle{IEEEbib}
\balance
\bibliography{paper}









\end{document}

%% file: body.tex
\begin{abstract}
    Inferring the quality of network services is the vital basis of optimization for network operators. However, prevailing real-time video streaming applications adopt encryption for security, leaving it a problem to extract Quality of Service (QoS) indicators of real-time video. In this paper, we propose DaI, a traffic-based real-time video quality estimator. DaI can partially decrypt the encrypted real-time video data and applies machine learning methods to estimate key objective Quality of Experience (QoE) metrics of real-time video. According to the experimental results, DaI can estimate objective QoE metrics with an average accuracy of 79\%. 
\end{abstract}

\section{Introduction}

Real-time video is playing a more and more important role in people's daily life, such as real-time video conferences, online collaborative work and real-time control. However, real-time video conferencing usually suffers from quality degradation caused by network fluctuations, such as video stalls, visual artifacts and other phenomena. To avoid or alleviate the quality degradation caused by these phenomena, network operators and real-time video service providers need to evaluate the Quality of Experience (QoE) of real-time video services and make further optimizations with network configurations and streaming strategies.

Network operators need to monitor the real-time video traffic and its quality transparently, but it is challenging when this traffic is encrypted. The challenge is twofold: \textbf{lack of transport layer features} and \textbf{encryption on the application layer}.




On the one hand, unreliable transport layer protocols provide fewer Quality of Service (QoS) indicators. 
Real-time video streaming adopts unreliable transport protocols such as User Datagram Protocol (UDP) because they are not forced to retransmit when packet loss happens and do not induce high recovery delay like reliable transport protocols such as Transport Control Protocol (TCP). However, UDP packets have shorter headers in their packets and do not incorporate the problems of delay, ACK/retransmission and other information, thus providing fewer QoS indicators. The traditional prediction methods of quality of experience (QoE) for Video-on-Demand (VoD) based on TCP and HTTP have been proven to be feasible, especially to predict some objective QoE metrics, such as video bitrate and stall~\cite{ManglaHAZ19, BronzinoSAMTF19, GuttermanGAWWKZ19, SeufertCWGL19, ShenZXZLD20, WuLCH21}. However, the success of these traditional methods is based on the premise that video traffic has abundant transport layer features. When it comes to real-time video streaming over UDP, it is challenging to estimate the quality of real-time video with few transport layer features.

On the other hand, the encryption in the application layer exacerbates the lack of features. When real-time video streaming adopts UDP, the key information about its error correction is contained in the application layer. Specifically, real-time video streaming usually adopts Forward Error Correction to achieve low-latency error correction, and this is an important feature of real-time video that can be used to estimate the quality of real-time video streaming~\cite{CarofiglioGLMPS21}. However, most prevailing real-time video conference platforms such as Zoom~\cite{zoom}, WebEx~\cite{webex} and Tencent Meeting~\cite{tencentmeeting} adopt encryption in the application layer for the sake of security, and the encryption in the application layer makes it difficult to extract features from this layer, which exacerbates the lack of features.

\begin{figure}
    \centering
    \includegraphics[width=0.96\linewidth]{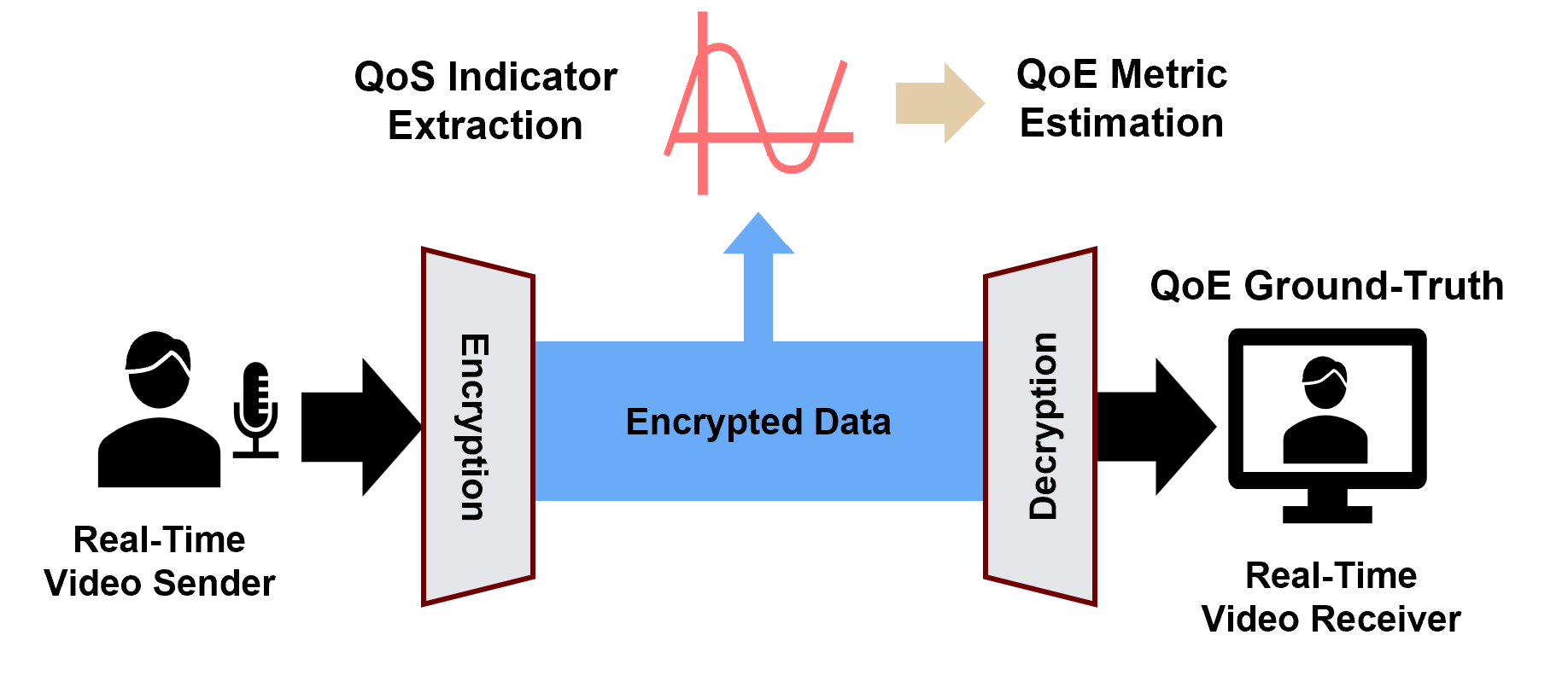}
    \caption{The overview of DaI. DaI can extract key QoS indicators from encrypted real-time video traffic and estimate objective QoE metrics.}
    \label{sys}
\end{figure}

In this paper, we present Decrypt and Infer (DaI), a real-time video stream quality inference system for network operators. As is shown in Fig.~\ref{sys}, DaI can decrypt and analyze the encrypted real-time video data stream, extract key QoS indicators such as video/FEC data rates and packet loss rate, and predict QoE metrics in real-time video streaming using machine learning methods. The QoE metrics include video resolution, frame rate and video bitrate. Taking Tencent Meeting~\cite{tencentmeeting} as the exemplary tested platform, DaI captures and analyzes the QoS indicators of Tencent Meeting's real-time video stream in a variety of semi-simulated network experiments for a long time, and estimates the QoE metrics of the real-time video stream. The comparison between the predicted value and the QoE ground truth recorded by the real-time video application proves that DaI can achieve an average accuracy of 79\%.

\section{Deciphering method \& QoS Extraction}

For the encrypted real-time video stream, we need to analyze the UDP data stream and extract the QoS features for the QoE prediction. We first introduce the real-time video stream quality indicators that can be obtained before decryption (Sec.~\ref{dec:before}), then introduce the decryption method of the encrypted video stream for the real-time video conference tested platform and describe the complete set of QoS features obtained after decryption (Sec.~\ref{dec:decipher}).

\subsection{Initial QoS metrics}
\label{dec:before}

\begin{table}
    \renewcommand\arraystretch{1.5}
    \centering
    \caption{Format of UDP Packet}
    \begin{tabular}{|c|c|}
        \hline
        16 bits source port & 16 bits destination port \\
        \hline
        16 bits datagram length & 16 bits checksum \\
        \hline 
        \multicolumn{2}{|c|}{payload}\\
        \hline
    \end{tabular}
    
    \label{tab:udp}
\end{table}

For the encrypted real-time video stream running over the UDP protocol, the payload of the UDP packet cannot be analyzed directly, and the initial QoS characteristics can only be obtained from the packet header of the UDP protocol. As is shown in the table.~\ref{tab:udp}, the header of the UDP packet contains limited information: the source port and the destination port can only be used to identify a certain UDP flow, and the checksum field is not useful for extracting QoS indicators. The \textbf{packet length} describes the size of the UDP packet and reflects the transmission rate of the entire UDP stream together with the timestamp of the packet. The \textbf{interval} between the arrival timestamps of the UDP packet sequence can also be used as a QoS feature. Therefore, the QoS features that can be directly extracted from the encrypted real-time video stream before decryption only include:
\begin{itemize}
    \item \textbf{Data rate of UDP stream}. It can be calculated by dividing the total byte length of the UDP payload by the length of the time interval for a given time interval.
    \item \textbf{Packet Arrival Interval}. It indicates the frequency of the UDP stream sending packets.
\end{itemize}

\subsection{Deciphering Method}
\label{dec:decipher}

\begin{table*}[t]
\begin{minipage}{\textwidth}
    \renewcommand\arraystretch{1.5}
\centering
\caption{Statistics of UDP payload bytes of encrypted real-time video stream I}
\begin{tabular}{|c|c|c|c|c|c|c|c|c|}

    \hline
    Byte position & 1 & 2 & 3 & 4 & 5 & 6 & 7 & 8  \\
    \hline
    Highest frequency byte value & \textbf{65} & 232 & 232 & 56 & 22 & 133 & 0 & 0  \\ 
    \hline
    Corresponding frequency & 99.5\% & 99.5\% & 35.1\% & 1.2\% & 0.6\% & 26.5\% & 0.9\% & 0.9\% \\
    \hline
    Byte position  & 9 & 10 & 11 & 12 & 13 & 14 & 15 & 16 \\
    \hline
    Highest frequency byte value & 10 & 196 & 186 & 134 & 175 & 69 & 138 & 27 \\ 
    \hline
    Corresponding frequency & 3.1\% & 99.5\% & 99.5\% & 68.8\% & 99.5\% & 99.5\% & 99.5\% & 99.5\% \\
    \hline
    \end{tabular}
    
    \label{dec:major1}
\end{minipage}
\\[12pt]
\begin{minipage}{\textwidth}
\centering
\renewcommand\arraystretch{1.5}
\caption{Statistics of UDP payload bytes of encrypted real-time video stream II}
\begin{tabular}{|c|c|c|c|c|c|c|c|c|}

    \hline
    Byte position & 1 & 2 & 3 & 4 & 5 & 6 & 7 & 8  \\
    \hline
    Highest frequency byte value & \textbf{65} & 131 & 74 & 247 & 70 & 174 & 791 & 245  \\ 
    \hline
    Corresponding frequency & 99.2\% & 95.6\% & 73.2\% & 3.9\% & 0.9\% & 9.1\% & 4.0\% & 4.0\% \\
    \hline
    Byte position  & 9 & 10 & 11 & 12 & 13 & 14 & 15 & 16 \\
    \hline
    Highest frequency byte value & 113 & 77 & 19 & 24 & 60 & 75 & 11 & 219 \\ 
    \hline
    Corresponding frequency & 4.2\% & 99.2\% & 95.6\% & 76.8\% & 95.6\% & 95.6\% & 95.6\% & 99.2\% \\
    \hline
    \end{tabular}
    
    \label{dec:major2}
\end{minipage}
\end{table*}

\subsubsection{Encryption Vulnerability}
Most of the encryption methods are based on bit exclusive or (XOR) operations. Let $M$ represent the information content, $K$ be the key, and $E$ be the encrypted message, then the encryption process can be expressed as follows:
$$E = M \oplus K$$
One-Time Pad (OTP) is a classic encryption and decryption method. This method requires the sender and the receiver to negotiate a common key before the message transmission, and the encryption and decryption processes are executed by XOR operations with the same key. The OTP encryption method cannot be effectively cracked when the key is completely random from the perspective of the hacker and each key is disposable, but these two points are not necessarily guaranteed in real use. Therefore, we first hypothesize that the real-time video stream encryption of the Tencent Meeting may not correctly use the encryption method, and verify our hypothesis through experiments. Considering the header format of traditional RTP (Real-time Transport Protocol), the unencrypted application layer packet header in the UDP payload in Tencent Meeting is likely to have the following RTP header properties:

\begin{property}
    Some fields in the header of the same real-time video stream are consistent. For example, they include the protocol version number, the stream identifier number SSRC (synchronization source) and some fixed configuration fields.

    Some fields in the header of the same real-time video stream show regular changes. For example, the sequence number may follow the law of continuous increase.
\label{dec:prop_1}

\end{property}

Therefore, if the key is not changed for some time during the transmission of a real-time video stream, the consistent fields in the application layer header are also consistent after XOR encryption. By calculating the occurrence frequency of the byte value of the head part of the UDP payload after encryption and observing the most frequent byte values, it can be judged whether the key is changed.

We did two simple experiments with Tencent Meeting in a stable network environment. We captured packets at one end through Wireshark software, identified the UDP stream that transmitted the real-time video conference through the total amount of transmitted data, and counted the frequencies of byte value on the first 16 bytes of UDP payload at the receiving end. Table.~\ref{dec:major1} and table.~\ref{dec:major2} record the byte values with the highest frequency and the corresponding frequency in the two experiments.

It is obvious that the UDP payload of the same real-time video stream on the tested platform has the following characteristics: 1) it starts with an unencrypted byte value of 65, which is the same between different real-time video streams; 2) There are extremely high-frequency byte values (higher than 99\%), and the highest frequency byte values of different groups of experiments are different, which indicates that it is likely that the fields are consistent before encryption, and the same encryption key is applied, resulting in the phenomenon that they are still consistent after encryption. \textbf{The experimental results prove that there is an encryption vulnerability in the tested platform: only a common key will be generated for a real-time video stream, and the key values at the same byte position of different packets are completely consistent.} This vulnerability makes it possible to decrypt (or partially) the real-time video stream of the tested platform and extract more QoS features.

\subsubsection{Distinguish between video data and FEC redundancy}
\label{dec:disting}
Considering the properties of the SSRC number and PT number in the traditional RTP protocol~\ref{dec:prop_2}, the UDP payload of the tested platform is also likely to fulfill the property before encryption. 
\begin{property}
    Video packets and FEC (forward error correction) packets belonging to the same real-time video stream have the same SSRC but have different payload types, which are identified by the PT (payload type) field.
\label{dec:prop_2}
\end{property}
Therefore, to distinguish whether the encrypted packets of the tested platform are video data and FEC redundant data, it is necessary to further identify the position of the PT field in the UDP payload of the tested platform.

Since real-time video streaming is usually equipped with a redundancy adaptive mechanism, i.e., the relative size of redundancy and video data will be adaptively adjusted according to the packet loss of the network, the proportion of data packets with PT of video type can be indirectly controlled by manually adjusting the packet loss of the network. At the same time, the highest frequency of each byte position of the UDP payload can be observed. So, we can manually control the network packet loss rate and observe the position of the highest byte frequency which is negatively correlated with the controlled packet loss rate.

\begin{figure}
    \centering
    \includegraphics[width=0.7\linewidth]{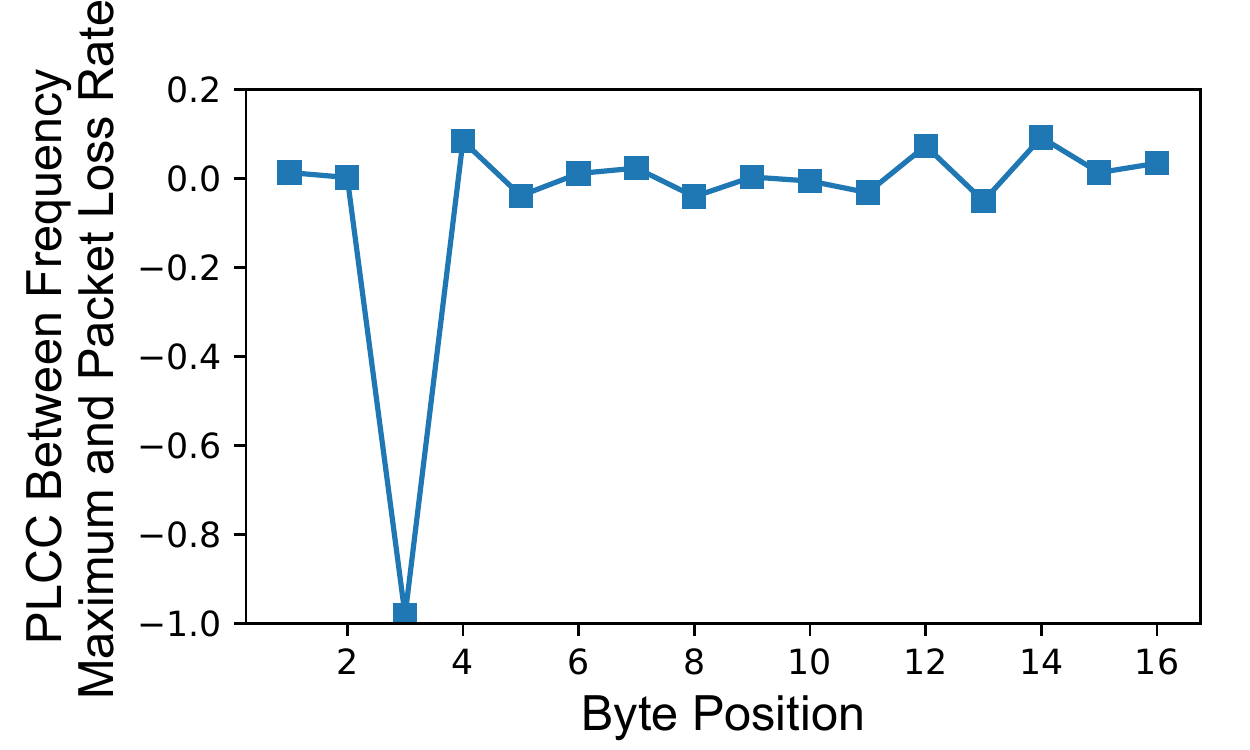}
    \caption{The linear correlation coefficients between the highest frequency of byte value and the packet loss rate in different byte positions in the UDP payload of the tested platform.}
    \label{chap4:ssrc}
\end{figure}

To manually adjust the network configuration, we let one end of Tencent Meeting join an 802.11n wireless network, and let the AP (access point) of the wireless network can access the external network through another wired network. In this way, the AP serves as a gateway to the Internet. Using the Linux TC program on the AP, we can manually set additional packet loss for the transmission network environment of the tested platform. The test network environment used here is consistent with Sec.~\ref{chap4:set}. Repeat the real-time meeting experiment for about 10 minutes under the condition that the preset additional uniform packet loss rates are 0\%, 2\%, 4\%, 6\%, 8\% and 10\% respectively. The selected experiment time is from 9:30 to 10:30 a.m. on weekdays to avoid severe congestion and fluctuations in the Internet. Calculating the linear correlation coefficient of the highest byte frequency and the packet loss rate at the first 16 bytes of the UDP payload of the tested platform, we get the figure.~\ref{chap4:ssrc}.

    
        

It is apparent from the figure.~\ref{chap4:ssrc} that the second byte of UDP payload conforms to the property.~\ref{dec:prop_2}, so it can be inferred that the highest byte frequency of this position can represent the proportion of video packets in the real-time video stream of the tested platform. In combination with the total transmission rate of the UDP stream, it is easy to obtain two important QoS indicators: \textbf{video throughput} and \textbf{FEC throughput} through multiplication.

\subsubsection{Extract packet loss of video stream}
Also considering the nature of the video packet sequence number in the traditional RTP protocol\ref{dec:prop_2}, the UDP payload of the tested platform is also likely to fulfill it before encryption.
\begin{property}
    The packet header of the video-type packet contains a sequence number field, and the sequence number field follows an increasing rule with an interval of 1 in one real-time video stream.
\label{dec:prop_3}
\end{property}
In section.~\ref{dec:disting}, we can already identify the data packets belonging to the video stream of the tested platform. Therefore, to extract the packet loss of the video stream of the tested platform, we need to locate the byte position of the sequence number field of the video data packet of the tested platform.

Unlike the analysis of the consistent value field, the positioning of the sequence number field is based on the new property.~\ref{dec:prop_4}.

\begin{property}
    The XOR result between two adjacent sequence numbers must be the power of $2$ minus one, and this property remains after the OTP encryption.
\begin{equation}
    \begin{split}
    n \oplus (n+1) = 2^p - 1\\
    n, p \in \mathcal{N}, p \ge 1
    \end{split}
\end{equation}
    
\label{dec:prop_4}
\end{property}

\begin{figure}
    \centering
    \vspace{0.1in}
    \includegraphics[width=0.7\linewidth]{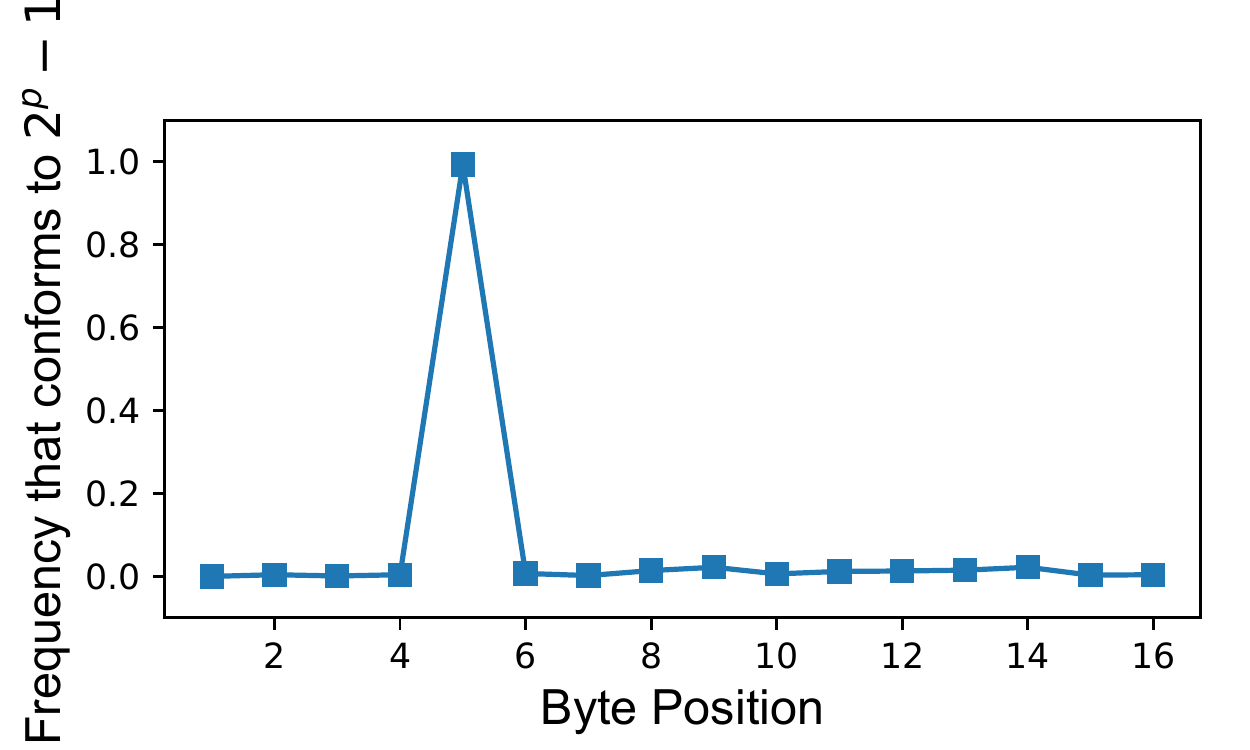}
    \caption{The value frequency of different byte positions in the UDP payload whose XOR result of adjacent video packets fulfills $2^p-1$ in the tested platform.}
    \label{chap4:loss}
\end{figure}

We carry out real-time video experiments in the experimental environment mentioned in section.~\ref{dec:disting} and extract XOR values of adjacent video packets on each byte position and conduct frequency analysis. In this experiment, we do not make any additional adjustments to the network configuration, to avoid packet loss as much as possible, which will break the property.~\ref{dec:prop_4} significantly. Figure.~\ref{chap4:loss} illustrates the frequency of adjacent packets of video stream at different byte positions whose XOR result fulfills the form of $2^p-1$.

As is shown in figure.~\ref{chap4:loss}, the fifth byte in the UDP payload of the tested platform perfectly fulfills the property.~\ref{dec:prop_4}. In addition, considering that the sequence number field may occupy multiple bytes, the fifth byte is the lowest byte of the big-endian integer. By extending the field, we finally found that the 4th to 5th bytes constitute a complete sequence number field. Therefore. It is easy to calculate the encryption key on this field during monitoring by analyzing the XOR field value of the adjacent packet. Then we can decrypt this field to extract sequence numbers and further calculate the \textbf{packet loss rate} of the video stream.

\section{QoE prediction method for real-time video stream}

This paper uses the random forest classification algorithm to predict the QoE metrics of the real-time video stream. Given a time window of a certain length, we could calculate the video throughput, FEC throughput, packet loss rate, and packet arrival time interval within the time window, and then predict the objective QoE metrics of the real-time video stream within the time window using the random forest classification algorithm. In the implementation, we let the length of the time window be 2 seconds. 

The objective QoE metrics of the tested platform considered in this paper include:
\begin{itemize}
    \item \textbf{video bitrate}. The encoding rate of the video is a classical objective QoE metric of video. A higher encoding rate usually produces a better user experience.
    \item \textbf{video framerate}. Generally speaking, the higher the frame rate, the smoother the video playback, and the easier the user can get a better experience.
    \item \textbf{video resolution}. The resolution also has a significant impact on the quality of the video. Too low resolution will directly lead to the blurring of the video picture, thus leading to a bad experience for the user.
\end{itemize}
The objective QoE metrics are divided into gears through discretization. The division method is shown in table.~\ref{chap4:qoesplit}.
\begin{table}
    \centering
    \renewcommand\arraystretch{1.5}
    \caption{Discretization of objective QoE metrics}
    \begin{tabular}{|c|c|c|c|c|}
        \hline
        QoE metrics$\backslash$Gears & High & Medium & Low & Very low \\
        \hline
        Bitrate/kbps & $\ge$700 & 500$\sim$700 & 300$\sim$500 & $\le$300 \\
        \hline
        Framerate/fps & $\ge$30 & 20$\sim$30 & 10$\sim$20 & $\le$10 \\
        \hline 
        Resolution/width & $\ge$1280 & 960$\sim$1280 & 640$\sim$960 & $\le$640 \\
        \hline
    \end{tabular}
    \label{chap4:qoesplit}
\end{table}
Since the real-time video of the tested platform generally maintains an aspect ratio of $16:9$, only the number of pixels on the width is considered when grading the video resolution.

\begin{figure}[t]
    \centering
    \vspace{0.1in}
    \includegraphics[width=0.96\linewidth]{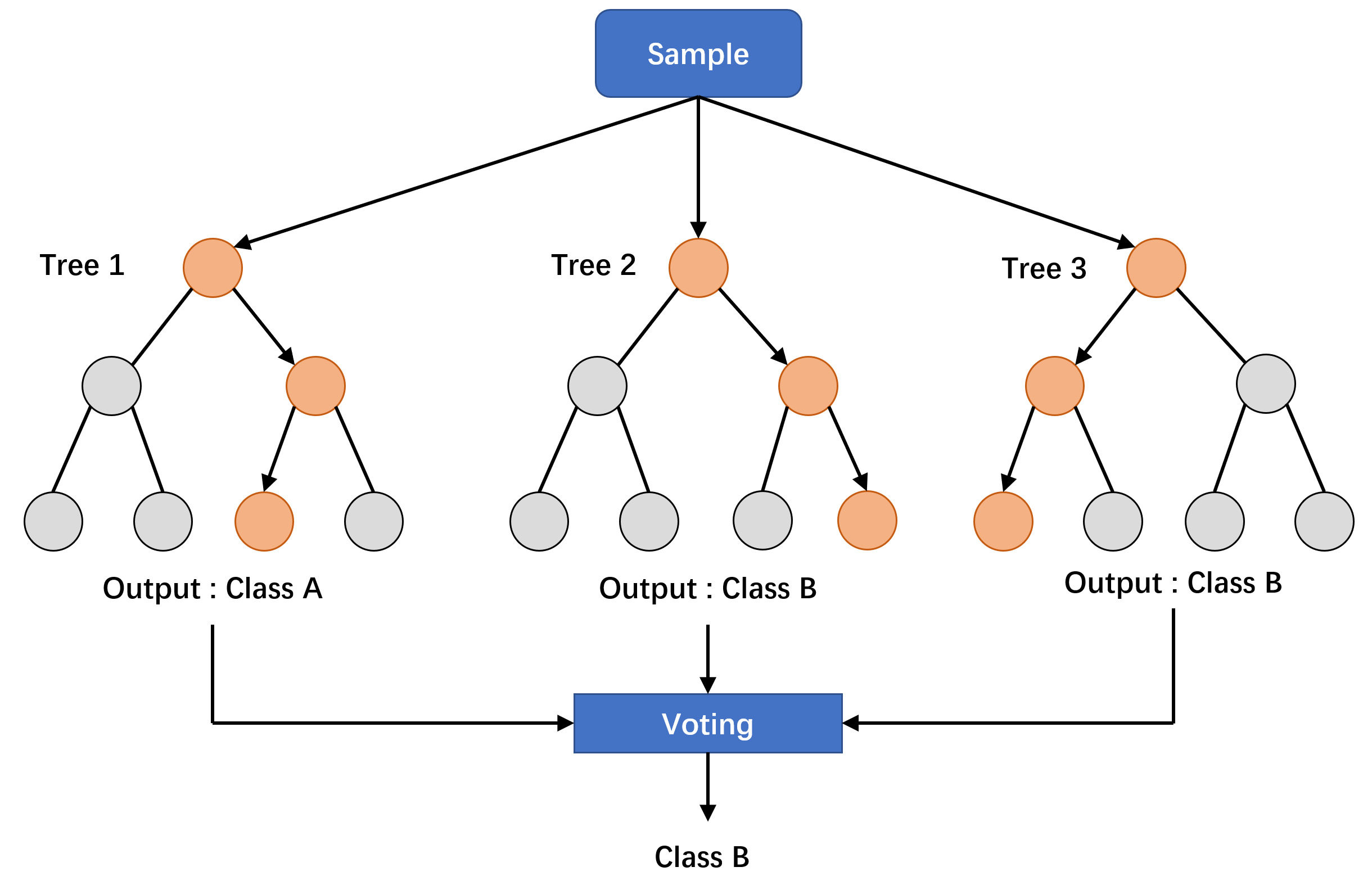}
    \caption{Schematic diagram of QoE metric prediction algorithm, random forest algorithm used by DaI.}
    \label{chap4:rf}
\end{figure}

Subsequently, DaI applies the random forest classification algorithm to build a classification model from the QoS indicators to the QoE metrics. DaI uses independent classification models for different prediction targets. Random forest is a tree-based machine learning algorithm, which combines the advantages of bagging and decision tree algorithms. It builds independent decision tree classifiers by selecting different sample subsets, and then determines the final output of the random forest learner according to the voting results of multiple decision tree classifiers. The schematic diagram of the random forest algorithm is shown in figure.~\ref{chap4:rf}.

\section{Experimental results and analysis}

\subsection{Experimental setup}
\label{chap4:set}
\begin{figure}[t]
    \centering
    \includegraphics[width=0.96\linewidth]{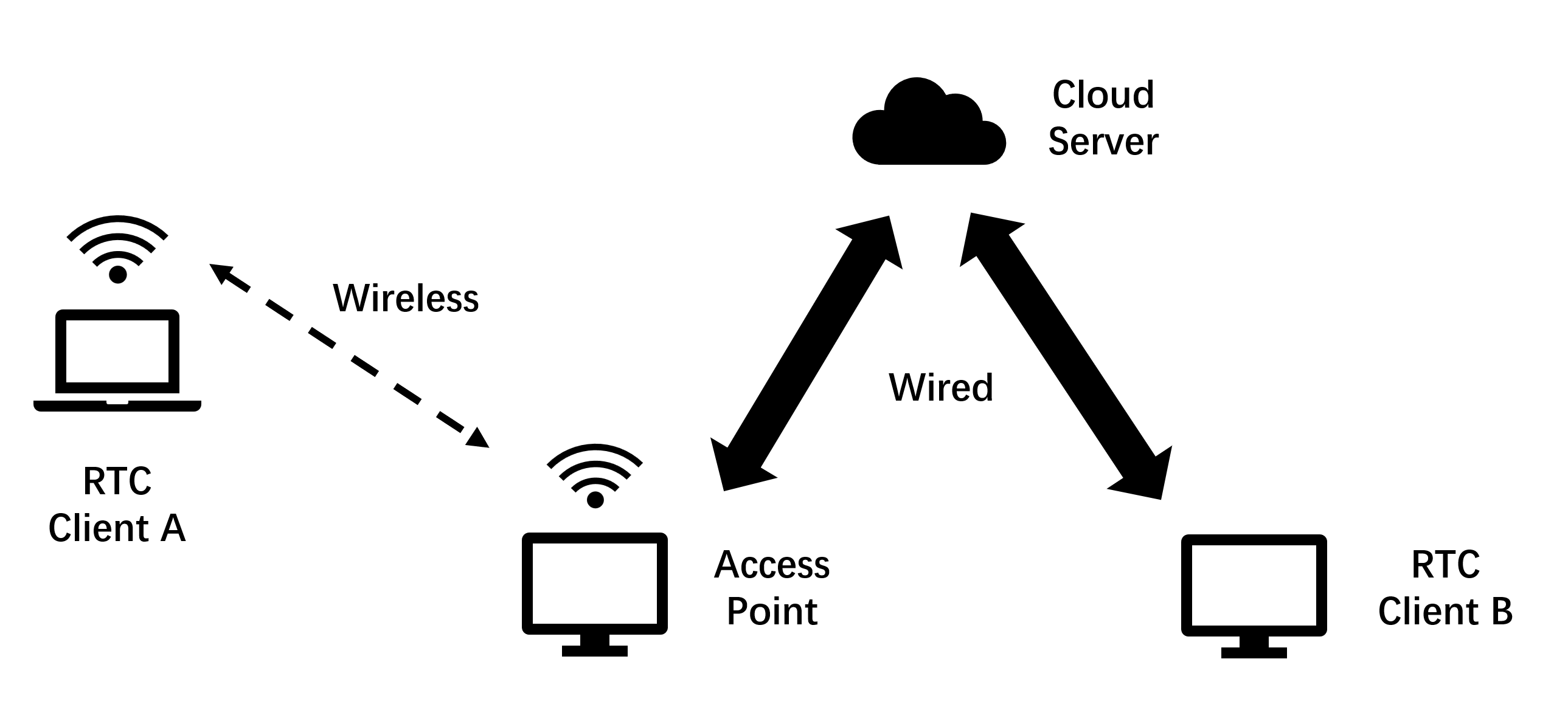}
    \caption{The topology of the tested network environment.}
    \label{chap4:expset}
\end{figure}

To prove that DaI is generally effective in monitoring and estimating the QoE of encrypted real-time video streams in different network environments, a semi-simulated test network environment is designed in this section, as shown in figure.~\ref{chap4:expset}. We focus on the real-time video stream transmitted by client B to client A and explore the relationship between the network QoS indicators monitored at the wireless network and the objective QoE metrics measured in client A. Using the Linux TC program on the intermediate router to modify the network conditions, the experiment can be carried out under some weak network conditions to some extent. The network condition modifications include:
\begin{align}
    \text{Maximum bandwidth} & \in \{900\text{kbps}, 1000\text{kbps}, 1100\text{kbps}\}\\
    \text{Extra packet loss rate} & \in \{0\%, 5\%, 10\%\}\\
    \text{Additional delay} & \in \{0\text{ms}, 100\text{ms}, 200\text{ms}\}
\end{align}
During the experiment, each combination of network condition modifications lasted for 1 hour, and a total of 27 hours of semi-simulation experiments were conducted. At the same time, by analyzing the log file of the real-time video application at client A, we can know the ground-truth value of the objective QoE metrics of client A at each time point.

\subsection{Experimental result}

\begin{figure*}[t]
    \centering
    \includegraphics[width=0.96\linewidth]{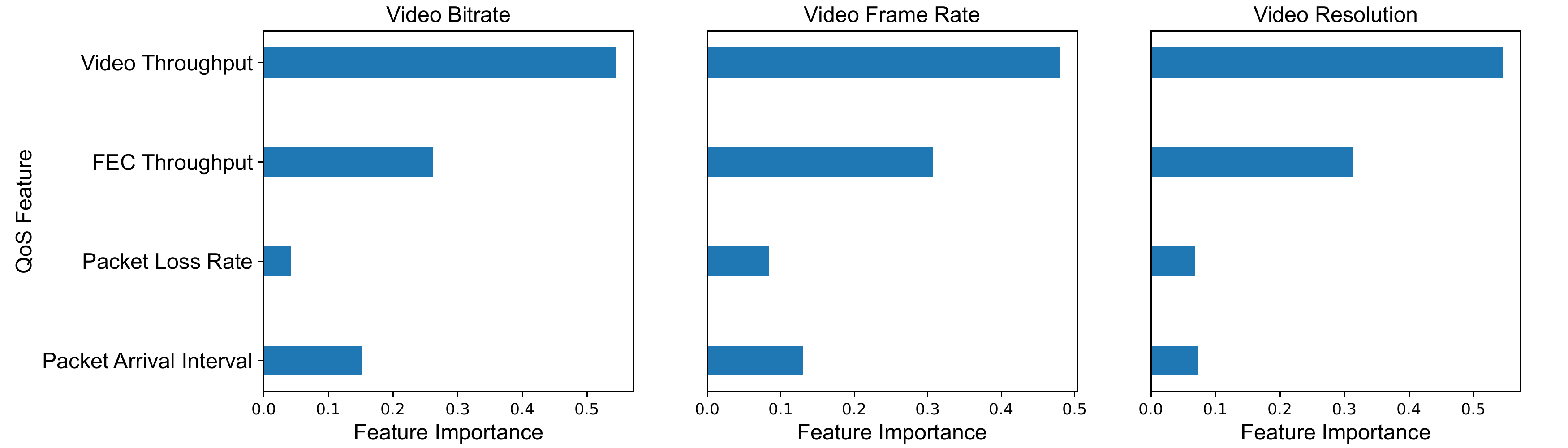}
    \caption{Feature importance of the prediction models for target objective QoE metrics.}
    \label{chap4:impor}
\end{figure*}

\textbf{Objective QoE prediction results}: By comparing the prediction result of DaI with the objective QoE metrics recorded by the application, the prediction accuracy of DaI can be obtained. Table.~\ref{chap4:r2score} records the prediction accuracy of each objective QoE metric, and figure.~\ref{chap4:impor} shows the distribution of feature importance of the prediction model of each objective QoE metric.

\begin{table}
    \centering
    \renewcommand\arraystretch{1.5}
    \caption{Accuracy of DaI in predicting various objective QoE metrics}
    \begin{tabular}{|c|c|c|}
        \hline
        Objective QoE metrics & Micro F1 Score (Accuracy) & Macro F1 Score \\ 
        \hline
        Video bitrate & 86.5\%  & 79.1\% \\
        \hline
        Video frame rate & 63.4\% & 51.5\% \\
        \hline
        Video resolution & 87.2\% & 46.7\% \\
        \hline  
    \end{tabular}
    \label{chap4:r2score}
\end{table}

From the table.~\ref{chap4:r2score}, it can be seen that DaI's accuracy of various objective QoE metrics of real-time video streams reached a high level, indicating that DaI can indeed monitor and estimate the quality of encrypted real-time video streams. Among the various objective QoE metrics, the prediction accuracy of the video bitrate and resolution is relatively high, while the prediction accuracy of the video frame rate is relatively low. This is probably due to the large adjustment range of the video frame rate at the sending end in the tested platform so there is a relatively large difference between the frame rate of the newly buffered video segments and the frame rate of the video segments being played when there is a certain video buffer at the receiving end. 

Judging from the importance of the characteristics of the prediction model, no matter which objective QoE metric is predicted, the video throughput and FEC throughput are absolutely important factors, which is consistent with our intuitive feeling: the video throughput is strongly related to the video encoding bitrate, and the FEC throughput will affect whether the real-time video can be played smoothly at the receiving end. However, the packet loss rate is not so important in all models. We guess that the fluctuation of the test network environment is not strong enough, so the packet loss is stable, and the ratio of the FEC throughput to the total throughput tends to be close to the packet loss rate, which makes the packet loss unable to provide more information.

\textbf{Equivalent prediction accuracy of subjective QoE}: After knowing the predictive performance of DaI for objective QoE metrics, it is natural to think of the question of whether DaI can further understand the subjective satisfaction of users. Therefore, we further explored the accuracy of DaI in predicting users' subjective QoE metrics. Considering that DaI directly outputs the prediction results of objective QoE metrics, we applied ITU P.1203 quality evaluation model.~\cite{raakegrlgf17} to further expand DaI and input the prediction results of DaI on objective QoE metrics into P.1203 model to obtain an estimated value of the user's subjective QoE. On the other hand, to obtain the true value of the comparative test, we applied P.1203, analyze and calculate the real-time video bitstreams saved by client A during the experiment, and obtain the "real" user subjective QoE results of the real-time video in the experiment. To measure the prediction accuracy of DaI, we selected the R2 score as the evaluation metric. For a set of real value $\{y_i \} ^ n $ and predicted value $\{y_i '\} ^ n $, R2 score is defined as follows:
\begin{equation}
    \text{R2 score} = 1 - \frac{\sum_{i=1}^n (y_i - y_i')^2}{\sum_{i=1}^n (y_i - \bar{y})^2}
\end{equation}
where $\bar{y} $is the average value of the set $\{y_i \} ^ n $. Table.~\ref{chap4:subj} describes the equivalent prediction accuracy of the subjective QoE of DaI.

\begin{table}
    \centering
    \renewcommand\arraystretch{1.5}
    \caption{The equivalent prediction accuracy of subjective QoE of DaI}
    \begin{tabular}{|c|c|c|}
        \hline
        Variance of subjective QoE ground-truth & 0.807 \\
        \hline
        Mean square error of subjective QoE prediction & 0.117 \\
        \hline
        R2 score & 0.85\\
        \hline  
    \end{tabular}
    \label{chap4:subj}
\end{table}

It can be seen from table.~\ref{chap4:subj} that the R2 score in the equivalent subjective QoE prediction of DaI has reached $0.85 $, which indicates that DaI has quite strong real-time video stream quality monitoring and prediction capability. Even used to predict the end-to-end subjective QoE metric, DaI can also maintain a high-performance level.

\section{Conclusion}

In this paper, we propose DaI, an encrypted real-time video quality monitoring system for network operators. DaI discovers and takes advantage of the vulnerability of the encrypted real-time video stream, obtains the QoS indicators of the real-time video stream by decrypting and extracting feature of the encrypted real-time video stream, and then used the machine learning method to predict the objective QoE metrics. The experiment in the semi-simulated network environment shows that DaI achieves high accuracy in predicting the objective QoE quality of the encrypted real-time video stream. In addition, extending DaI and predicting the subjective QoE metric can also obtain brilliant performance. Therefore, DaI is a very potential encrypted real-time video stream quality monitoring and prediction system.